\documentclass[prb,twocolumn,showpacs]{revtex4}
\usepackage{amsmath,graphics,latexsym}

\begin{document}

\title{Ballistic Spin Injection and Detection in
Fe/Semiconductor/Fe Junctions}
\author{Phivos Mavropoulos, Olaf Wunnicke, and Peter H. Dederichs}
\affiliation{Institut f\"ur Festk\"orperforschung, Forschungszentrum
J\"ulich, D-52425 J\"ulich, Germany}
\date{\today}

\begin{abstract}
  We present {\it ab initio} calculations of the spin-dependent
  electronic transport in Fe/GaAs/Fe and Fe/ZnSe/Fe (001) junctions
  simulating the situation of a spin-injection experiment. We follow a
  ballistic Landauer-B\"uttiker approach for the calculation of the
  spin-dependent dc conductance in the linear-responce regime, in the
  limit of zero temperature. We show that the bulk band structure of
  the leads and of the semiconductor, and even more the electronic
  structure of a clean and abrupt interface, are responsible for a
  current polarisation and a magnetoresistance ratio of almost the
  ideal 100\%, if the transport is ballistic. In particular we study
  the significance of the transmission resonances caused by the
  presence of two interfaces.
\end{abstract}

\maketitle

\section{Introduction}

The controlled spin-dependent electronic transport through
magnetic/nonmagnetic heterostructures is a central issue in the rising
field of spin electronics.\cite{Prinz98,Wolf01} In some cases, such as spin
valves or giant magnetoresistance devices, the basic-science
discoveries have lead to technological applications within less than a
decade. In other cases, however, as {\it e.g.}~spin injection into the
conduction band of semiconductors (SC), much remains yet to be
understood and achieved, experimentally and theoretically.

The interest in spin injection from a ferromagnetic (FM) material into
a semiconductor has been largely motivated by the proposed, but not
yet achieved, spin field-effect transistor of Datta and
Das.\cite{Datta90} There have been many tries, with increasing
success, to demonstrate that such a device is
feasible.\cite{Kikkawa99,Malajovich01,Fiederling99,Ohno99,Schmidt01,Hammar99,Monzon99,Gardelis99,Filip00,LaBella01,Zhu01}
It has been already shown\cite{Kikkawa99,Malajovich01} that electrons
in the conduction band of semiconductors can travel long distances
without loosing memory of their spin. In parallel, many attempts to
achieve spin-polarised currents have been made. The use of magnetic
semiconductors as leads of the
junction\cite{Fiederling99,Ohno99,Schmidt01} would be a possibility,
but they have the drawback of low Curie temperature, and thus would
not be applicable at room-temperature. On the other hand, the attempts
to use metallic ferromagnetic contacts were at first non-promising.
Efforts to use InAs-based contacts\cite{Monzon99,Gardelis99,Filip00}
due to their nice properties of an abrupt interface and an ohmic
transition have resulted in very low current polarisation, which might
sometimes even be attributed to stray-field Hall or magnetoresistance
effects.\cite{Tang00} Several theoretical approaches based on the
spin-diffusion or the Boltzmann equation have shed light on the
behaviour of such systems.\cite{several,Schmidt00,Rashba00,Fert01}
Recently Schmidt {\it et al.}\cite{Schmidt00} revealed a basic
obstacle for succesful spin injection, namely the conductivity
mismatch between the FM and the SC, resulting in too low current
polarisation unless the FM contact is almost 100\% spin polarised.
Their conclusion holds in the diffusive regime, when one can use a
resistor model for the FM/SC/FM heterostructure. To overcome this
fundamental difficulty, Rashba\cite{Rashba00} and Fert and
Jaffres\cite{Fert01} have proposed that the FM and SC parts should be
separated by a tunneling spin-polarising slab, the high resistance of
which would balance the drawback of the conductivity mismatch of a
direct contact. In parallel, and independently from these suggestions,
there has been the observation of Grundler\cite{Grundler01} that a
ballistic transistor would allow for a higher current polarisation
than a diffusive one, and that this should be realisable if a
two-dimensional electron gas was used.  Already, simple model
calculations,\cite{Grundler01,Hu01,Heersche01} based on a
free-electron approach of the electronic structure of the leads, have
shown that ballistic transport can give spin injection efficiencies of
a few percent.  Much more is seen, though, when one takes into account
the full band structure of the FM material and the self-consistent
electronic structure of the interface. Indeed, as first proposed by
Kirczenow,\cite{Kirczenow01} one can have ideal spin filters if the FM
Fermi surface of only the one spin direction, when projected to the
plane of the interface, has no states in the part of the
two-dimensional Brillouin zone where the conduction band starts, so
that there is no propagation into the SC from this spin channel. This
``selection rule'' unfortunately does not apply in certain interesting
systems such as Fe/GaAs or Fe/ZnSe. Nevertheless, as shown by Wunnicke
{\it et al.},\cite{Wunnicke02} in these systems the interface
reflectance is so much different for the two spin directions that one
gets spin injection ratios as high as 99\% in an {\it ab initio}
ballistic calculation. Apart from the theoretical efforts, there are
some very encouraging recent experiments of Zhu {\it et
  al.}\cite{Zhu01} giving already a few percent of current
polarisation in Fe/GaAs(001).

In the current article we present {\it ab initio} calculations of
ballistic spin-dependent transport in Fe/GaAs/Fe and Fe/ZnSe/Fe
trilayer heterostructures grown epitaxially in the $<$001$>$ direction
emulating a spin-valve geometry. In this way we extend the work of
Wunnicke {\it et al.}\cite{Wunnicke02} to include spin injection {\it
  and} detection.  We show that the presence of the two spin-filtering
interfaces increases the current polarisation even closer to the ideal
100\%, and we also calculate the high magnetoresistance ratios of
these structures, which is also approaching the ideal 100\%. We
observe interesting interference effects due to the presence of two
interfaces, and give an aspect of the whole problem that brings it in
close connection with the theory of magnetic tunnel junctions as it is
described in Refs.~\onlinecite{MacLaren99} and
\onlinecite{Mavropoulos00}. Our results thus stress that epitaxial
junctions operating as close as possible to the ballistic regime can
form almost ideal spin filters and can exhibit extremely high
magnetoresistance ratios.

The article is organised as follows: in Section~\ref{sec2} we present
the basic formulae of our {\it ab initio} approach. In
Section~\ref{sec3} we describe the junctions to be calculated and the
approximations made. The role of the symmetry of the wavefunctions in
transport through Fe/SC/Fe junctions is explained in
Section~\ref{sec4}. Sections~\ref{sec5} and \ref{sec6} contain the
results for the current spin polarisation and a discussion of
interesting interference resonance effects, while Section~\ref{sec7}
is devoted on the case of antiparallel orientation of the leads and
the magnetoresistance properties. Finally we discuss the limitations
of our approach and conclude with a summary in Section~\ref{sec8}.

\section{Method of calculation \label{sec2}}

Our calculations are based on density-functional theory in the local
spin density approximation (LDA). We employ the screened
Korringa-Kohn-Rostoker (KKR) Green's function method\cite{Zeller95} to
calculate the electronic structure of the systems. In this
multiple-scattering approach, the one-electron retarded Green's
function at energy $E$ is written in terms of local wavefunctions
$R_L^n(\mathbf{r})$ and $H_L^{n}(\mathbf{r})$ (regular and irregular
solutions of the single-site Schr\"odinger equation, respectively,
characterised by the angular momentum index $L=(l,m)$), centered at
lattice sites $\mathbf{R}_n$ and $\mathbf{R}_{n'}$, as
\begin{eqnarray}
\lefteqn{G(\mathbf{R}_n+\mathbf{r},\mathbf{R}_{n'}+\mathbf{r}')}&& \nonumber\\
&=&-i\sqrt{E}\sum_L R_L^n(\mathbf{r}_{<}) H_L^{n'}(\mathbf{r}_{>}) 
\delta_{nn'}\nonumber\\
&&+ \sum_{LL'} R_L^n(\mathbf{r}) G_{LL'}^{nn'}(E) R_{L'}^{n'}(\mathbf{r}')
\label{eqKKR1}
\end{eqnarray}
with $G_{LL'}^{nn'}(E)$ the so-called structural Green's function
describing the intersite propagation; $\mathbf{r}_{<}$ and
$\mathbf{r}_{>}$ are respectively the shorter and longer of
$\mathbf{r}$ and $\mathbf{r}'$, and atomic units have been used
($e=-\sqrt{2}$, $\hbar=1$, $m=1/2$). The structural Green's function
is related in turn to the known Green's function of a reference system
via an algebraic Dyson equation. For more details on this we refer the
reader to Refs.~\onlinecite{Papanikolaou02} and \onlinecite{Wildberger97}.

The systems consist of two half-infinite (Fe) leads, assumed to have
perfect periodicity otherwise. Sandwiched between these leads is an
``interaction'' region where a different material (SC) can be placed
and where the scattering of the Bloch waves takes place. The
interaction region and the two leads have common in-plane Bravais
vectors, {\it i.e.}~in-plane ($x$-$y$) periodicity (perpendicular to
the growth direction). If needed, larger (non-primitive)
two-dimensional unit cells are taken to match the lattice constants of
the materials. The two-dimensional periodicity of the layered systems
allows to Fourier-transform the Green's function in the $x$ and $y$
directions, obtaining a two-dimensional Bloch vector
$\mathbf{k}_{\parallel}=(k_x,k_y)$ as a good quantum number, and
retaining an index $i$ to characterise the layer in the direction of
growth $z$. The Green's function connecting the layers $i$ in the left
lead and $i'$ in the right lead is then written
\begin{eqnarray}
\lefteqn{
G(\mathbf{R}_i+\chi_{\nu}+\mathbf{r},\mathbf{R}_{i'}+\chi_{\nu'}+\mathbf{r}')
}&&
\nonumber\\
&=&\frac{1}{4\pi^2S_{\mathrm{SBZ}}}
\int_{\mathrm{SBZ}} d^2k_{\parallel} \,
e^{i\mathbf{k}_{\parallel}(\chi_{\nu}-\chi_{\nu'})} \nonumber\\
&&\times\sum_{LL'} R_L^i(\mathbf{r})
G_{LL'}^{ii'}(\mathbf{k}_{\parallel};E) R_{L'}^{i'}(\mathbf{r}')
\label{eqKKR2}
\end{eqnarray}
where $\mathbf{\chi_{\nu}}$ and $\chi_{\nu'}$ are in-plane lattice
vectors, $\mathbf{R}_i$ is the interlayer lattice vector, SBZ is the
surface Brillouin zone of the system and $S_{\mathrm{SBZ}}$ its area.
In this equation each layer $i$ is assumed to have a unique atom type,
hence only the index $i$ suffices to characterise the local
wavefunction. In the case of more inequevalent atoms per layer, an
extra index is introduced to account for the propagation between
different kinds of atoms. Moreover, in the case of ferromagnetism, the
Green's function is different for each spin direction
$\sigma=\uparrow$ or $\downarrow$.

For the calculation of the conductance in linear response all the
information needed is contained in the Green's function. In the
Landauer-B\"uttiker approach,\cite{Landauer57,Buttiker86} which
identifies the ballistic conductance $g$ with the transmission
probability of the conducting channels, one has
\begin{equation}
g = \frac{e^2}{2\pi\hbar}\sum_{\sigma,\mathbf{k}_{\parallel}}\sum_{\mu,\mu'}
T(\mathbf{k}_{\parallel},\mu,\mu',\sigma)
\label{eqLandauer}
\end{equation}
relating the transmission probability $T$ per channel to the
conductance $g$. Here each channel is characterised by the band index
$\mu$, the $\mathbf{k}_{\parallel}$-vector and the spin $\sigma$ of
the incoming electrons, and similarly by the primed indices for the
outgoing electrons, both having the same Fermi energy $E_F$.
Conservation of spin due to assumed absence of spin-orbit scattering,
and of $\mathbf{k}_{\parallel}$ due to two-dimensional periodicity,
have allowed us to omit the summation over $\sigma'$ and
$\mathbf{k}_{\parallel}'$ in the outgoing electron channels.  We
follow here the formalism of Baranger and Stone,\cite{Baranger89}
relating $g$ to the spatial derivative of the Green's function
connecting a cross-sectional plane in the left lead ($L$) to one in
the right lead ($R$). It is assumed that these planes lie in the
asymptotic regime, where interface perturbations and evanescent
interface states are no longer present. The formula for the
$\mathbf{k}_{\parallel}$-projected conductance
$g(\mathbf{k}_{\parallel},\sigma)$ per two-dimensional unit cell
surface area and spin $\sigma$ reads
\begin{widetext}
\begin{equation}
g(\mathbf{k}_{\parallel},\sigma) = -\frac{1}{4\pi^3} \int_L d^2r \int_R
d^2r' G_{\sigma}(\mathbf{r},\mathbf{r}';\mathbf{k}_{\parallel};E_F)
\overleftrightarrow{\partial_z} \overleftrightarrow{\partial_{z'}}
G_{\sigma}^*(\mathbf{r},\mathbf{r}';\mathbf{k}_{\parallel};E_F)
\label{eqBaranger1}
\end{equation}
\end{widetext}
where the symbol $\overleftrightarrow{\partial_z}$ stands for
\begin{equation}
f(\mathbf{r})\overleftrightarrow{\partial_z} g(\mathbf{r})=
f(\mathbf{r}){\partial_z} g(\mathbf{r})-({\partial_z} f(\mathbf{r}))
g(\mathbf{r}). 
\end{equation}
The conductance is evaluated only at the Fermi level $E_F$ since we
are at the limit of zero temperature. The complex conjugation in the
last term of eq.~(\ref{eqBaranger1}) comes from conversion of the
advanced Green's function to the retarded one by conjugation and
exchange of $\mathbf{r}$ and $\mathbf{r}'$.
$G(\mathbf{r},\mathbf{r}';\mathbf{k}_{\parallel})$ is given by the
two-dimensional Fourier transform of (\ref{eqKKR2}). By virtue of the
Fourier transformation, the integration in (\ref{eqBaranger1}) is not
performed over the whole lead cross-sectional area, but only over a
two-dimensional unit cell. The total conductance per two-dimensional
unit cell surface area for each spin channel is then
\begin{equation}
g_{\sigma} = \int_{\mathrm{SBZ}} d^2k_{\parallel} \,
g(\mathbf{k}_{\parallel},\sigma).
\end{equation}
Current conservation guarantees that the result is independent of the
position of the cross-sectional planes of integration, as long as they
are chosen in the asymptotic region. Details about the evaluation of
the conductance will be given elsewhere.\cite{Mavropoulos02}

The formula we use for the conductance has be proven to be equivalent
to the Landauer-B\"uttiker formula.\cite{Baranger89} The conductance
we calculate is then fully ballistic; diffuse scattering is assumed to
be absent.  Our approach also ignores spin-orbit scattering and any
spin-flip events. We must also note that the semiconductor band
gaps are known to be underestimated in the LDA by a factor of about
50\%. This can have some quantitative significance, but the trends of
our results are expected remain unaltered even if we choose to enlarge
the gap artificially.

In the calculations, the atomic sphere approximation (ASA) for the
potentials is used, {\it i.e.}~they are assumed to be spherically
symmetric around each atomic site and to occupy an atomic volume; on
the other hand, the full charge density, rather than its spherically
symmetric part, is taken into account. Moreover, we treat the systems
nonrelativistically. An angular momentum cutoff of
$l_{\mathrm{max}}=2$ has been taken for the wavefunctions and Green's
funcions in the self-consistency procedure.

\section{The systems under study \label{sec3}}

We study the spin-dependent transport through Fe/GaAs/Fe and
Fe/ZnSe/Fe junctions. The junctions are supposed to have grown
epitaxially on (001) interfaces, and in an ideal way so that the
transition from one material to another is abrupt. Absence of
interdiffusion and of disorder are assumed; in this way, we are
dealing with a system grown in the $z$ direction and being
translationally invariant in the $x$ and $y$ directions . The Fe leads
are supposed to be infinite, while the semiconductor thickness is
varied from 41 to 97 monolayers (ML). In such thicknesses, the
evanescent interface states in the semiconductor are expected to have
decayed to insignificance compared with the Bloch wavefunctions, so
the transport will be mediated through propagating states. 

Throughout the system, the experimental Fe lattice constant of
$a_{\mathrm{Fe}}=2.871$\AA\ is used. Thus, all atoms sit on ideal
positions of an underlying bcc lattice. In particular, in the SC part,
the zincblende structure can be easily seen to fit on such a lattice,
with half of the bcc sites occupied by Zn and Se (or Ga and As) atoms
and the rest occupied by vacancies. Viewed in this way the consecutive
positions of the atoms in the cubic diagonal of the bcc lattice are
(Zn, Se, vacancy, vacancy). The zincblende lattice constant is then
twice the one of the bcc. One can see, that using $2\times
a_{\mathrm{Fe}}=5.742$\AA\ in the SC part results only in a slight
mismatch of less than 2\%, the experimental lattice constants being
5.654\AA\ for GaAs and 5.670\AA\ for ZnSe. In all cases, Zn
termination of the ZnSe spacer and Ga termination of the GaAs spacer
was considered. As shown in Ref.~\onlinecite{Wunnicke02}, the spin
polarisation of the current through the single interface for the other
terminations (Se and As) is also extremely high, and from the analysis
of Sections~\ref{sec5}-\ref{sec7} it follows that the two-interface
junctions for those terminations will have qualitatively the same
properties as the ones studied here. The two planes $L$ and $R$ used
for the integration were 6MLs away from the interfaces in the Fe
region, where the asymptotic regime is assumed to have been reached.
Variation of this distance causes insignificant changes in the
results.

In a system as the ones we are considering, the Fermi level will be
naturally determined by the infinitely long Fe leads. But in the
spacer material, two or three monolayers after the interface, the
potentials and the charge density must be almost bulk-like. For this
reason, the potentials of the inner atoms of the spacer will be
automatically adjusted to the Fe Fermi level by a constant shift which
is the result of the interface dipole layer. The self-consistent
calculation of the potential close to the interface is then essential.

Since we want to inject electrons into the SC conduction band, we must
emulate in some way a gate voltage, or energy shift, acting on the SC
potentials in order to lower the conduction band minimum slightly
under the Fermi level. This artificial shift is different than the one
just mentioned above, and it enters as a parameter in our
calculations. We avoid disturbing the interface electronic structure,
which is strongly influenced by the metal-induced gap states, and
proceed as follows.\cite{Wunnicke02} The first two SC monolayers
adjacent to the interface are kept as calculated by a self-consistent
calculation of a 9ML-thick SC slab sandwiched between infinite Fe
leads. The same applies also for the first neighbouring Fe MLs. Having
saved the interface in this way, we take for the rest of the SC spacer
(third up to last-but-two ML) the bulk-like potential that we find for
the atoms in the middle of this Fe/9ML SC/Fe junction. This is
justified, since it is known that the potential stabilises quickly as
mentioned previously. The emulation of the gate voltage is achieved by
applying to this potential an extra shift such that the conduction band
minimum $E_c$ of this bulk-like structure falls slightly under the
Fermi level $E_F$ of the whole structure:
\begin{equation}
E_F = E_c + E_0.
\end{equation}

The parameter $E_0$, characterising the assumed gate voltage, is
varied in our calculations over three values: 20mRy, 10mRy, and 5mRy
(272meV, 136meV and 68meV, respectively). In this way we are able to
view the approach to small values as a limiting procedure; as we shall
see, these values are already in the limit of large spin polarisation
of the current and magnetoresistance.  Viewing the semiconductor part,
the small values of $E_0$ mean that the energy dispersion relation is
nearly parabolic,
\begin{equation}
E(\mathbf{k}) - E_c \simeq
\frac{1}{m^*} \mathbf{k}^2 = 
\frac{1}{m^*} (\mathbf{k}_{\parallel}^2 + k_z^2) 
\label{eqdisp1}
\end{equation}
where $m^*=(\partial^2 E/\partial k^2)^{-1}$ is the effective mass,
and that the Fermi wavenumber $k_F$ is very small:
\begin{equation}
E_F - E_c = E_0=\frac{1}{m^*} k_F^2 
= \frac{1}{m^*}(\mathbf{k}_{\parallel}^2 + k_z^2).
\label{eqdisp2}
\end{equation}
These relations are relevant in the semiconductors considered here
because of their direct band gap at the $\Gamma$-point. Because $E_0$
is very small, we have a very small Fermi sphere in the semiconductor.
For this reason, very few channels $\mathbf{k}_{\parallel}$ will be
able to conduct, namely those close to the center of the Brillouin
zone with $|\mathbf{k}_{\parallel}|\leq k_F$. For the rest $k_z$
becomes imaginary and represents decaying wavefunctions. These can
give rise to a tunneling current, but for the larger spacer
thicknesses they are small compared to the contribution from the
central part of the Brillouin zone. In any case they are always
included in the calculation.

\section{The importance of symmetry \label{sec4}}

As mentioned in the previous section, we are expecting contributions to
the current only from the central part of the (001) Surface Brillouin
Zone (SBZ), {\it i.e.}~from $\mathbf{k}_{\parallel}$ close to the
$\bar{\Gamma}$-point. In view of this, we will examine the expected
behaviour for states exactly at $\mathbf{k}_{\parallel}=0$, and argue,
and show in fact in the calculations, that by continuity the close-by
states will behave similarly.

To begin with, we must clarify that the two-dimensional unit cell and
the SBZ are determined by the SC part, since one SC lattice constant
is assumed to match exactly two Fe lattice constants in our case. The
states with $\mathbf{k}_{\parallel}=0$ can be examined in a great
extent through their symmetry properties, since the $z$-axis remains
invariant under many point-group operations. The single Fe (001)
surface is characterised by the symmetry group $C_{4v}$, having eight
operations: a fourfold rotation axis (here the $z$-axis) plus
reflections over the planes containing the $z$-axis and the $xy$
diagonal or antidiagonal.  But the zincblende (001) surface has the
symmetry group $C_{2v}$, having four operations (a twofold rotation
axis plus the reflections over the $xy$ diagonal and antidiagonal),
and being a subgroup of the former. As a result, the combined Fe/SC
interface is characterised by the group $C_{2v}$.

The idea now, in view the Landauer approach and
Eq.~(\ref{eqLandauer}), is to investigate the incoming states at the
Fermi level deep in the Fe lead, as incoming channels, in order to see
if their symmetry properties allow them to couple to SC propagating
bulk states, and then see if these in turn are allowed (by symmetry)
to couple to the outgoing states which propagate deep in the other Fe
lead. The different character of the Fe states for majority and
minority spin will give us in this way hints about the spin
polarisation of the current. This procedure can be used to propose
theoretically ideal spin filter systems. But note that in this way we
can only find which channels are excluded from transmission by
symmetry.  As we shall see, some channels can be almost blocked for
other reasons, contributing (by their absence from transmission) to
the spin injection effect.

\begin{figure}
  \begin{center}
    \leavevmode
    \rotatebox{270}{\resizebox{!}{8.5cm}{
        \includegraphics{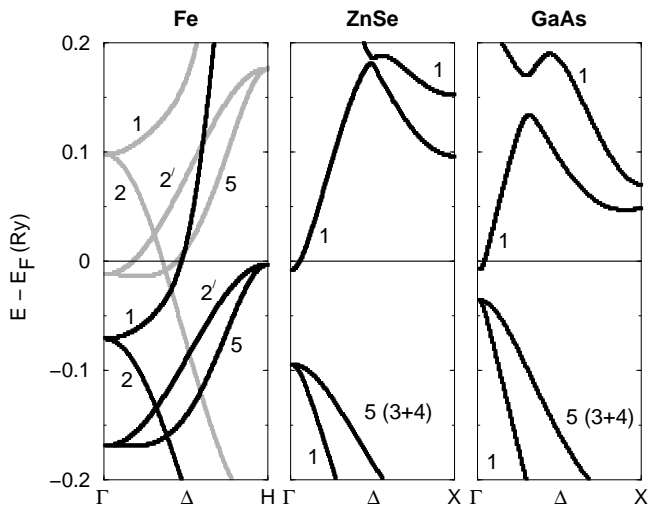}
        }}
    \caption{Energy bands of bulk Fe (left) together with bulk ZnSe
      (centre) and bulk GaAs (right) along the $\Delta$-direction
      ($k_z$), corresponding to $\Gamma-\mathrm{H}$ in bcc (Fe) and to
      $\Gamma-\mathrm{X}$ in fcc (zincblende). For Fe, the black lines
      represent majority-spin states, and the gray lines minority-spin
      states. The potentials of GaAs and ZnSe have been appropriately
      shifted so that the Fermi level falls slightly in the conduction
      band. Each band is named by the corresponding irreducible
      representation of the point group; {\it e.g.}~1 means the
      $\Delta_1$ representation, $2'$ the $\Delta_{2'}$ etc. For the
      notation see, {\it e.g.},
      Ref.~\protect{\onlinecite{Parmenter55}}. Note that the
      $k_z$-axes at the semiconductor plots should actually be half
      the size shown, since the lattice constant is assumed double the
      one of Fe. Backfolded bands due to the doubling of the Fe
      two-dimensional unit cell are unimportant and not shown.}
    \label{figFeGaAsZnSebands}
  \end{center}
\end{figure}
We can now turn our attention to Fig.~\ref{figFeGaAsZnSebands}, where
the energy bands of Fe, ZnSe and GaAs are drawn for
$k_x=k_y=0$ in the $k_z$-direction, which is the one of
interest as discussed earlier.  Each of them is named by the
irreducible representation to which it belongs\cite{Parmenter55} for
rotations around the $\Delta$-axis ({\it i.e.}~$k_z$). For example,
the state labeled ``1'' corresponds to the $\Delta_1$, which means
that the states are invariant under all group operations (rotations
around the $z$-axis); the label $2'$ refers to the $\Delta_{2'}$
representation, being invariant under reflections from the planes
containing the $z$-axis and either the $xy$ diagonal or antidiagonal.
But we must note that for Fe the nomenclature refers to the $C_{4v}$
group, while the symmetry group of the whole system as well as of the
bulk semiconductor is $C_{2v}$.  Therefore we must use the
compatibility relations between the two groups, that show us which
representations of $C_{4v}$ have nonzero projection in each
representation of $C_{2v}$. These can be found, for instance, in
Ref.~\onlinecite{Parmenter55}. In our case we see that, at the Fermi
level, only one band exists in the semiconductor (both for GaAs and
ZnSe), and it belongs to the representation $\Delta_1 (C_{2v})$ (in
parentheses we specify the point group to which the representation
belongs). With this representation, only the $\Delta_1 (C_{4v})$ and
$\Delta_{2'} (C_{4v})$ states of bulk Fe are compatible.  This means
that incident states of only these symmetries can couple to the
semiconductor conduction states (or even to each other, near the
interface) and propagate into the SC spacer, while the rest, $\Delta_2
(C_{4v})$ and $\Delta_5 (C_{4v})$, are totally reflected at the
interface.

Now, the energy bands of Fe can are quite different for majority
vs.~minority electrons near the Fermi level, due to the exchange
splitting. At $E_F$ the majority electrons have a $\Delta_1
(C_{4v})$-state that can couple to the semiconductor, while this is
absent for the minority-spin carriers. For these, on the other hand, a
$\Delta_{2'} (C_{4v})$-band exists that can do the job.  We note in
passing that this absence of $\Delta_1 (C_{4v})$ is due to the
so-called $s$-$d$-hybridisation gap which splits this band in two and
which happens to fall around $E_F$ for the minority-spin states.

If the $\Delta_{2'} (C_{4v})$-band were absent, or if it could not
couple to the $\Delta_1 (C_{2v})$-band of the semiconductor, we would
be facing an ideal spin filter: only majority-spin would be able to
propagate. Even in our case, however, we shall see that almost ideal
spin filtering will occur, because the two kinds of states, $\Delta_1
(C_{4v})$ and $\Delta_{2'} (C_{4v})$, have very different transmission
probabilities through the interfaces such that the $\Delta_{2'}
(C_{4v})$ channels are nearly blocked.

\section{Results for the spin-dependent conductance \label{sec5}}

\begin{figure}
\begin{center}
\resizebox{8.5cm}{!}{
\includegraphics{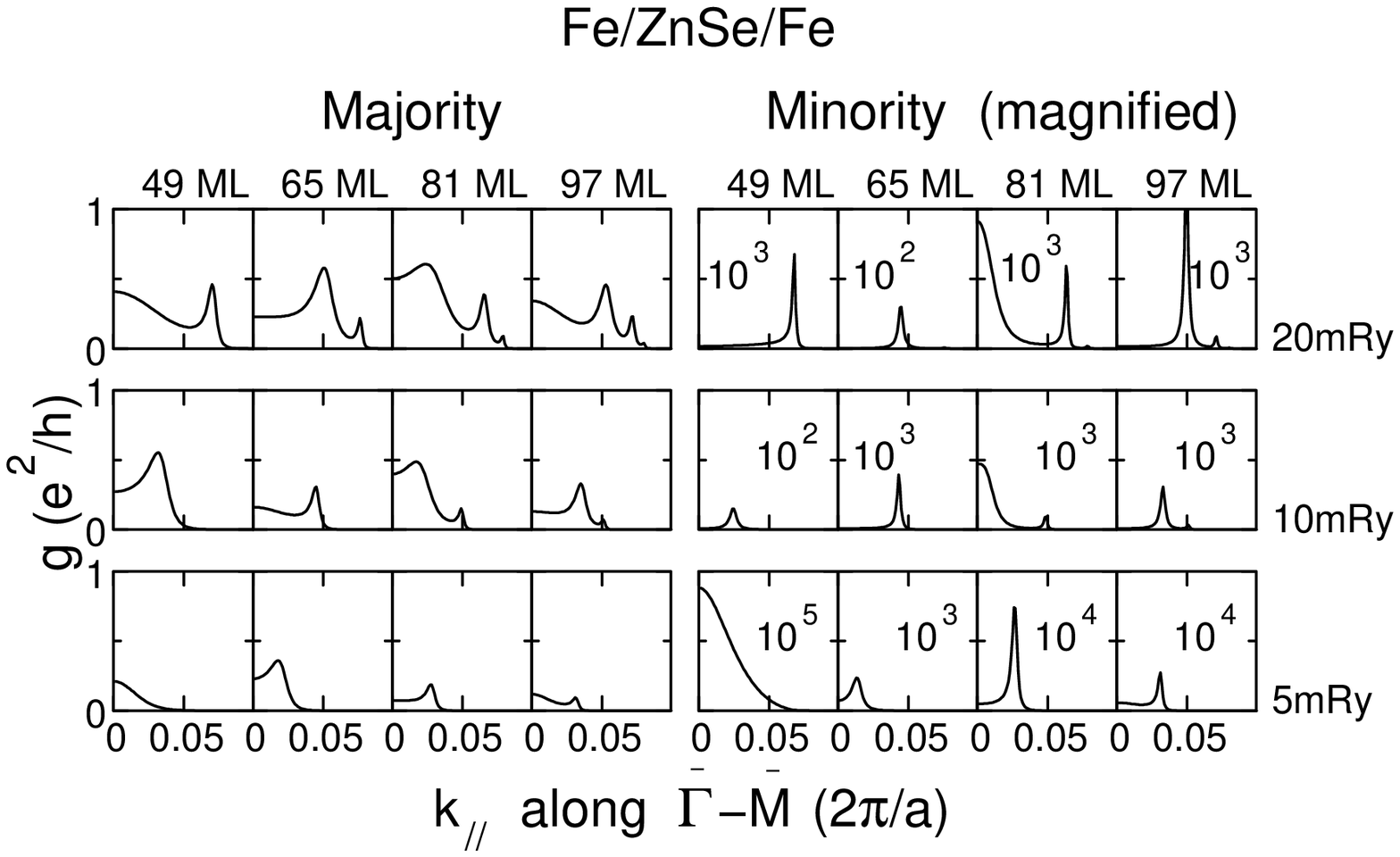}
}
\resizebox{8.5cm}{!}{
\includegraphics{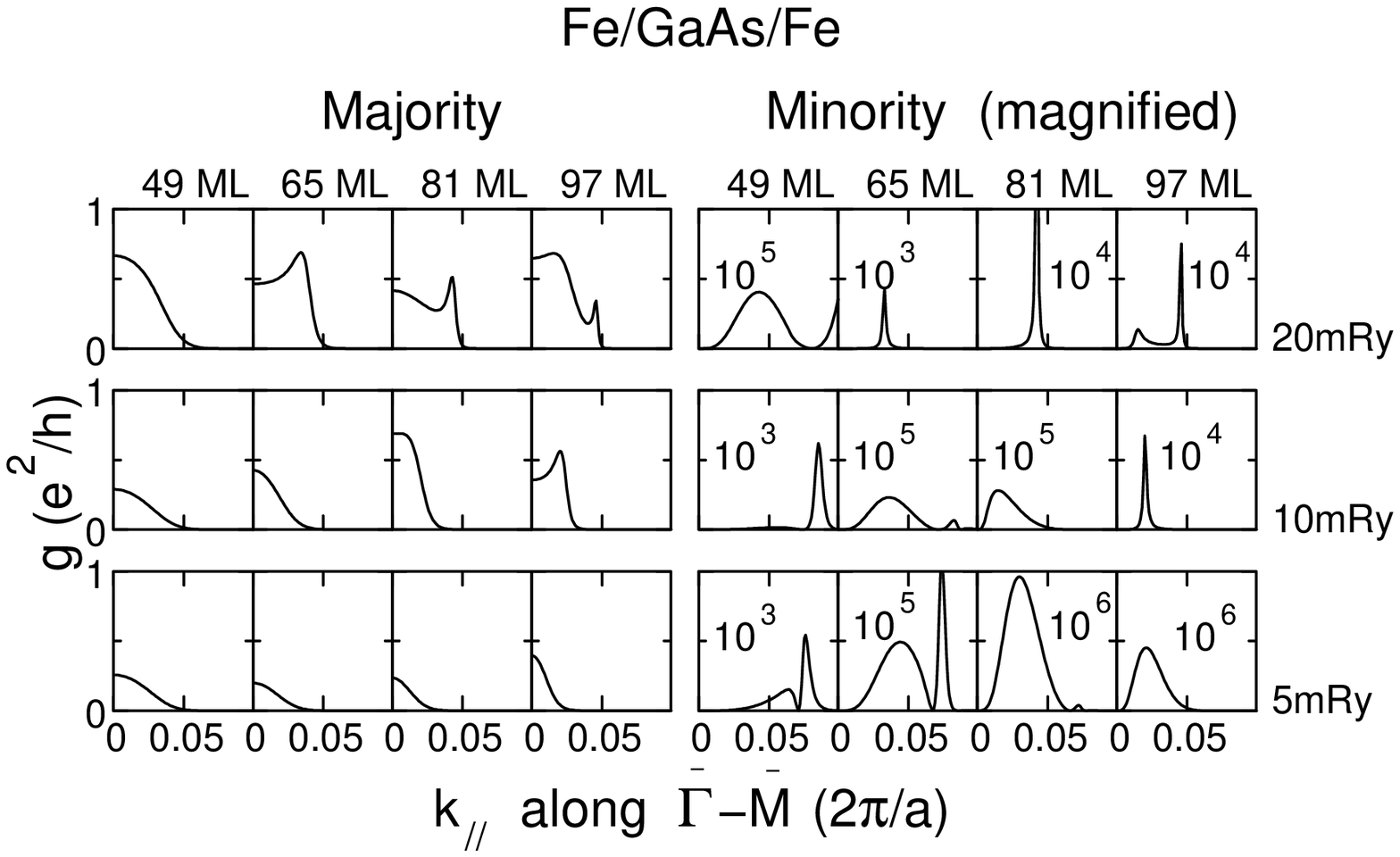}
}
\end{center}
\caption{Spin-dependent conductance for Fe/ZnSe/Fe (top) and
  Fe/GaAs/Fe (bottom) junctions, as a function of $k_x$, for the
  parallel magnetic configuration of the leads. The majority-spin
  conductance is illustrated in the left pannels and the minority in
  the right. Several SC spacer thicknesses are considered (49ML to
  97ML), and gate energy shifts of $E_0=5$, 10, and 20mRy. For the
  minority-spin case a magnification from $10^2$ to $10^6$ (see inset
  numbers) has been used to bring the graphs to the same scale. The
  $k_F$-values for ZnSe are 0.038, 0.056, and 0.083, and for GaAs
  0.021, 0.031, and 0.050 for $E_0=5$, 10, and 20mRy, respectively.}
\label{figZnSeGaAscond}
\end{figure}

The spin-dependent conductance as a function of
$\mathbf{k}_{\parallel}$ for several spacer thicknesses is shown in
Fig.~\ref{figZnSeGaAscond} for GaAs and ZnSe spacers, and for several
energy shifts $E_0$. The wavevector $\mathbf{k}_{\parallel}$ has been
taken along the $\Gamma$-$\mathrm{X}$ cubic direction, which in the
two-dimensional geometry corresponds to
$\bar{\Gamma}$-$\bar{\mathrm{M}}$. It is most convenient to express
$\mathbf{k}_{\parallel}$ in units of $2\pi/a_{\mathrm{SC}}=2\pi/(2
a_{\mathrm{Fe}})$, since this corresponds to the two-dimensional
periodicity of the whole system; henceforth these units will be
implied but omited for simplicity. The calculated values of $k_F$ are
given in the caption of Fig.~\ref{figZnSeGaAscond}.

The first evident observation is that the conductance practically
vanishes for $k_{\parallel}>k_F$, as expected. This effect shows up
clearer for the thicker spacers. Thus we can see that, as $E_0$ rises
and the Fermi sphere in the SC becomes larger, the cutoff in
conductance moves to higher values of $k_{\parallel}$, exactly as
$k_F$. As mentioned earlier, for larger values only evanescent states
can exist, giving rise to a very small tunneling current which dies
out as the spacer gets thicker. Nevertheless our calculations show
that these states dominate the behaviour in the small thickness
region. 

One can clearly see that the minority-spin conductance is lower by
orders of magnitude than the majority counterpart. This is clearly the
effect of the Fe minority $\Delta_{2'}$-state not being able to couple
well with the SC $\Delta_1$-state at the interface. The reason for
this is that the $\Delta_{2'}$-state consists locally of $d_{xy}$-like
site-centered orbitals. These point in-plane and are quite localised,
so they cannot overlap very well with the SC $\Delta_1(C_{2v})$
orbitals.  Moreover, the SC $\Delta_1(C_{2v})$ band consists of $s$,
$p_z$, and $d_{xy}$-like states. The latter are in fact the ones that
do couple to the $\Delta_{2'}$ minority band of Fe.  But we must note
that such $d_{xy}$-like SC states are not inherent to the SC atoms,
but rather induced as a distortion to the inherent $sp$ SC orbitals by
the neighbouring atoms sitting in the tetrahedral positions and giving
a directional preference; in this sence they appear just as a
correction when we use an angular momentum basis. As we depart from
the $\bar{\Gamma}$-point, other Fe minority orbitals (the
continuations of the $\Delta_5$ and $\Delta_2$ bands) begin to couple
slowly, so the transmission increases.

In contrast, the $\Delta_1(C_{4v})$-band present in the majority-spin
states consists locally of $d_{z^2}$, as well as $s$ and $p_z$-like
atomic orbitals; these, pointing partly into the SC and being more
extended, favour a better overlap and bonding with the SC states. Thus
the reflectance of the interface is by far stronger for the
minority-spin electrons, and a strongly polarised current results.

The arguments presented here show that one needs a clean and abrupt
interface, so that $\mathbf{k}_{\parallel}$ is conserved. In the case
of $\mathbf{k}_{\parallel}$-violation due to diffuse scattering the
effect of spin selection will be reduced. Indeed, the total ({\it
  i.e.}~$\mathbf{k}_{\parallel}$-integrated) density of states of Fe
at $E_F$ is higher for the minority-spin than for the majority-spin.
On these grounds one would expect even a negative current
polarisation, in similarity with Julliere's model\cite{Julliere75} for
spin-dependent tunneling; this might be the case if strong diffusive
scattering intermixes the scattering $\mathbf{k}_{\parallel}$-channels
in a completely random way.  Thus it is the specific selection rule
imposed by the interface in the ballistic regime that causes the
strong positive current polarisation.  We may also note that for other
interfaces, such as (110) or (111), the symmetry of the various
incident states is different than in (001), and the selection rule
might not be as strong; an {\it ab initio} calculation is necessary in
order to judge this.

Note that the same effect appears when one looks at tunneling, rather
than spin injection, in these structures. This is demonstrated in
Ref.~\onlinecite{MacLaren99}, where the tunneling through the
semiconductor is also confined to the states close to $\bar{\Gamma}$;
there, the majority Fe state of symmetry $\Delta_1$ couples much
better at the interface than the minority state of symmetry
$\Delta_{2'}$, while both propagate with equal difficulty afterwards
in the SC, as evident by the equal decay rate. In that case, of
course, one must consider the complex band structure of ZnSe in the
gap region as the analytical continuation of the conduction band of
the same symmetry for the interpretation of the
effect,\cite{Mavropoulos00} rather than the real conduction band
structure, but both tunneling and spin injection can be viewed in this
respect in a unified way.

\section{Interface reflectance and quantum well states \label{sec6}}

Another interesting feature is the multi-peaked structure of
$g(k_{\parallel})$. This is an interference effect to the discussion
of which we turn now. We start with the observation that the presence
of two Fe/SC interfaces can give rise to interference effects due to
the coherent multiple reflection of the electrons between them. So,
one expects resonances in the transmission, similarly to the case of a
square barrier of finite length met by free electrons of energy higher
than the barrier.\cite{Schiff55} More concretely, let us assume that
the transmission through each Fe/SC interface (1 or 2) has an
amplitude $t_{1,2}$ and the reflection $r_{1,2}$. These contain the
phase shifts $\phi_{1,2}$, that the wavefunction obtains for each
reflection, plus a phase factor of $e^{ik_z D}$ for the wave
propagation from side to side of the SC slab of thickness $D$ leading
to a phase of $2k_z D$ for a come-and-go. A resonance in transmission
will be formed whenever there is constructive interference after a
number of comes-and-goes of the wave; {\it i.e.}~one has to sum up the
series
\begin{eqnarray}
  t_{\mathrm{tot}} &=& t_1 t_2 + t_1 r_2 r_1 t_2 + t_1 r_2 r_1 r_2 r_1
  t_2 + \cdots   \nonumber \\
 &=& t_1 \frac{1}{1-|r_1| |r_2| e^{i(2k_zD + \phi_1 +
      \phi_2)}} t_2 
\label{eqmultref}
\end{eqnarray}
in order to find the maxima in transmission. If the two interfaces are
the same, as in Fe/SC/Fe with parallel magnetic orientation of the
leads, the single-interface probabilities $T_{\mathrm{si}}$ of
transmission and $R_{\mathrm{si}}=1-T_{\mathrm{si}}$ of
reflection are equal for the two interfaces (the Fe/SC interface is
equally hard to cross in either direction), and by squaring the
previous equation one finds the total transmission probability to be
\begin{eqnarray}
T_{\mathrm{tot}} &=& |t_{\mathrm{tot}}|^2   \\ 
&=&\frac{T_{\mathrm{si}}^2}{1+(1-T_{\mathrm{si}})^2-2(1-T_{\mathrm{si}})
\cos(2k_zD+\phi_1+\phi_2)}\nonumber 
\label{eqres}
\end{eqnarray}
where we have used the fact that $T_{\mathrm{si}}=|t_1 t_2|$ is
valid in this case; Eq.~(\ref{eqres}) is equivalent to the formula of
Airy for a Fabry-Perot interferometer.  This function is clearly
oscillatory in $k_zD$, and it exhibits a maximum of
$T_{\mathrm{tot}}=1$, {\it i.e.}~a resonance, whenever the condition
for constructive interference is met:
\begin{equation}
\phi_1+\phi_2+2 k_z D=2\pi n,
\label{eq2}
\end{equation}
with $n$ an integer.

For a given thickness $D$, variation of $k_{\parallel}$ will cause
variation of $k_z$, and this will lead to these resonance
phenomena.\footnote{Similar interference resonances in
  $g(\mathbf{k}_{\parallel})$ have been observed in calculations of
  tunneling junctions,\cite{Butler01} for transmission through
  those evanescent states that also have a nonzero real $k_z$. Of
  course there the transmission probability is always many orders of
  magnitude smaller than one.} This is realised by combining
eqs.~(\ref{eqdisp2}) and (\ref{eqres}), so that the multi-peaked
structure in Fig.~\ref{figZnSeGaAscond} is
explained. To see what one expects qualitatively, we combine
eqs.~(\ref{eqdisp2}) and (\ref{eq2}) to get
\begin{equation}
2\,\sqrt{k_F^2-k_{\parallel}^2}\, D=2\pi n+\phi_1+\phi_2
\label{eq5}
\end{equation}
as a resonance condition. For zinc-blende structures, where $k$ varies
between $0$ and $1$ in units of $2\pi/a_{\mathrm{SC}}$, and
$a_{\mathrm{SC}}=4\mathrm{ML}$ in the (001) direction, the condition
relates $k_{\parallel}$ to the number of monolayers $N_{\mathrm{ML}}$:
\begin{equation}
\sqrt{k_F^2-k_{\parallel}^2}\, N_{\mathrm{ML}} = 2n+(\phi_1+\phi_2)/\pi.
\label{eq6}
\end{equation}
Naturally, $\phi_1$ and $\phi_2$ depend on $\mathbf{k}_{\parallel}$.
This formula can be seen to give three resonances already for
$N_{\mathrm{ML}}=100$ and $k_F=0.05(2\pi/a_{\mathrm{SC}})$.

Between the maxima there are minima of
$T_{\mathrm{tot}}=T_{\mathrm{si}}^2/(2-T_{\mathrm{si}})^2$.
For low values of $T_{\mathrm{si}}$ the halfwidth of the resonance
becomes very small; this is reflected at the minority-spin conductance
where the resonances are much more narrow and peaked, with extremely
low valued valleys between them, and thus their
$\mathbf{k}_{\parallel}$-integrated contribution remains insignificant
compared to the majority one. These arguments also demonstrate that
the interference effects are in practice unable to invert the injected
current polarisation, in contrast to what has been predicted by recent
model calculations.\footnote{Recent theoretical
  articles\cite{Mireles01,Schaepers01} based on model calculations
  have suggested that a modulation of the current polarisation due to
  interference by variation of $k_F$ \emph{is} possible. Our arguments
  deny this possibility for the systems considered here, but surely
  not for \emph{all} systems. Note that Refs.~\onlinecite{Mireles01}
  and \onlinecite{Schaepers01} refer to systems of lower
  dimensionality (2D or 1D), where the averaging over
  $\mathbf{k}_{\parallel}$ has a lesser effect (or is nonexistent, in
  1D). Also, the aforementioned modulation would be perhaps possible
  also in 3D if the minority-spin reflectance of the interface were
  lower.} We also note that, for $E_F \rightarrow E_C$,
$T_{\mathrm{si}}\propto k_z \propto \sqrt{E_F-E_C}$.\footnote{The
  situation is analogous to the one-dimensional case of a free
  electron of wavenumber $k_{\mathrm{in}}=\sqrt{E}$ encountering an
  infinitely long step barrier of height $V_0$ where the wavenumber is
  $k_{\mathrm{out}}=\sqrt{E-V_0}$.  The transmition probability there
  is $T_{\mathrm{si}}=4k_{\mathrm{out}}k_{\mathrm{in}}/
  (k_{\mathrm{out}}+k_{\mathrm{in}})^2$ going to zero as
  $\sqrt{E-V_0}$ for small $k_{\mathrm{out}}$. The point is that the
  group velocity in the barrier region goes to zero, although the
  transmission amplitude $t$ remains finite, so current conservation
  forces $T_{\mathrm{si}}=|t|^2 k_{\mathrm{out}}/k_{\mathrm{in}}$
  also to go to zero. The same point is true in our case when we
  approach the conduction band edge,\cite{Wunnicke02} and thus the
  analogy is valid.} Then for a given spacer thickness
$T_{\mathrm{tot}}$ goes to zero linearly as $T_{\mathrm{tot}}\propto
E_F-E_C$, while the first resonance appears for a thickness
$D_{\mathrm{res}}$ increasing to infinity as $1/\sqrt{E_F-E_C}$. In
the model described here, one can readily substitute the values of
$T_{\mathrm{si}}$ from a single-interface
calculation,\cite{Wunnicke02} and get the correct trend.\footnote{This
  applies also for the $k_{\parallel}$-dependence of $g$: one sees the
  multi-resonant form of Fig.~\ref{figZnSeGaAscond} if one substitutes
  the value for $T_{\mathrm{si}}(k_{\parallel}=0)$ taken from the
  single interface calculation into eq.~(\ref{eqres}), and takes for
  the other $\mathbf{k}_{\parallel}$'s
  $T_{\mathrm{si}}(\mathbf{k}_{\parallel})=T_{\mathrm{si}}(0)k_z/k_F
  = T_{\mathrm{si}}(0)\sqrt{k_F^2-k_{\parallel}^2}/k_F$; in such
  an approximation, of course, any anisotropy of
  $T_{\mathrm{si}}(\mathbf{k}_{\parallel})$ is lost.}
Nevertheless, in the calculations we cannot observe a perfect
resonance of transmission one, because perfect coherence is destroyed
by a very small but nonzero imaginary part of the energy, numerically
necessary for the calculation of the retarded Green's
function.\footnote{This small imaginary part $\epsilon$, in principle
  infinitesimal, can also be viewed as imposing an attenuation
  $\exp(-\sqrt{\epsilon}D)$ on the waves crossing a spacer of
  thickness D. Its effect can be studied by replacing $t_2\rightarrow
  t_2 \exp(-\sqrt{\epsilon}D)$ and $r_2\rightarrow r_2
  \exp(-2\sqrt{\epsilon}D)$ in eq.~(\ref{eqmultref}). It turns out
  that the resonance maxima are then reduced for larger thicknesses,
  and that the reduction is stronger for smaller
  $T_{\mathrm{si}}$. Thus the minority-spin conduction maxima are
  much more affected.}

\begin{figure}
\begin{center}
\resizebox{0.5\textwidth}{!}{
\includegraphics{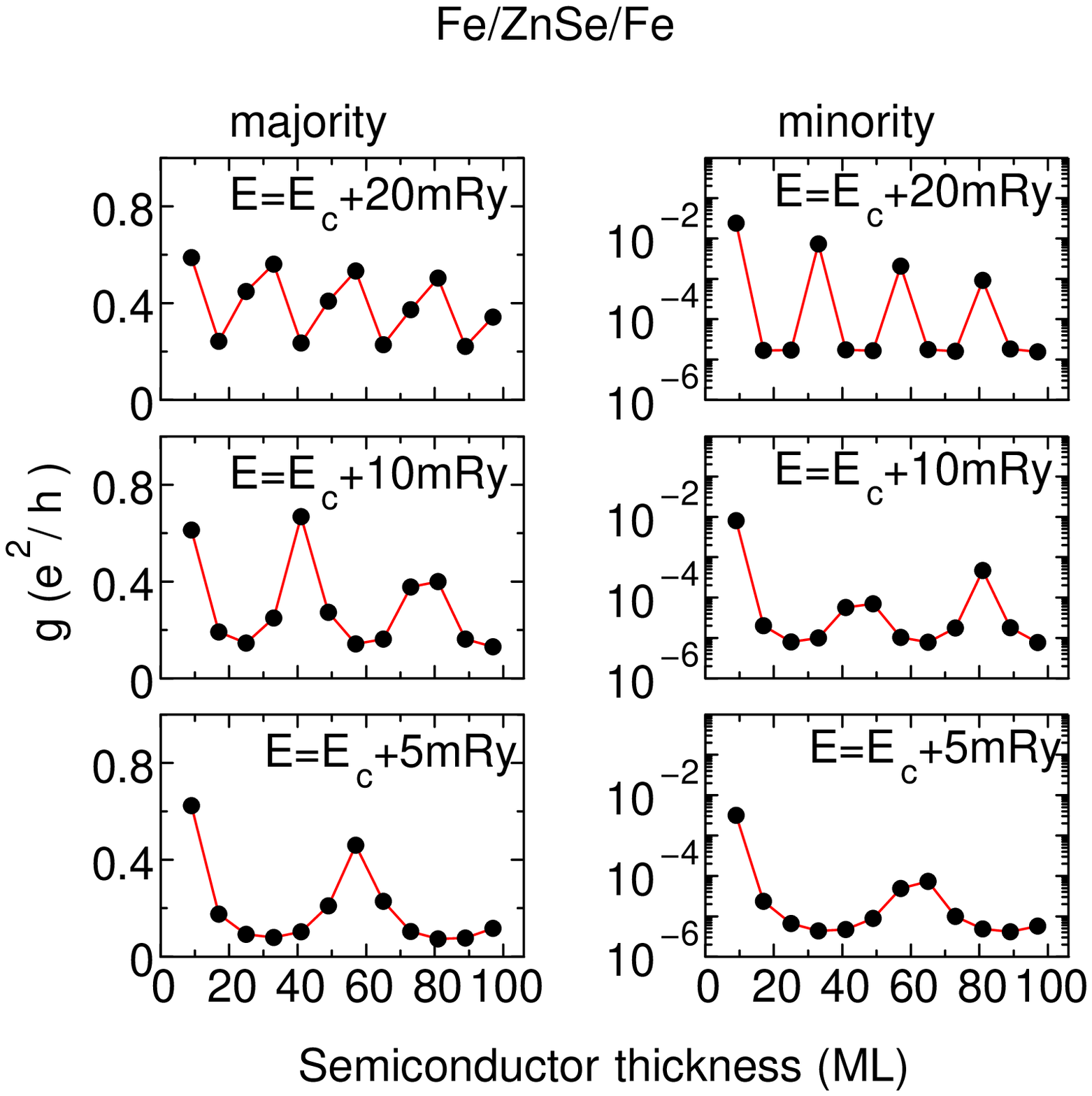}}
\end{center}
\caption{Majority- (left) and minority- (right) spin conductance at
  $\mathbf{k}_{\parallel}=0$ as a function of the ZnSe spacer
  thickness. The oscillations of period $2\pi/(2k_F)$ are evident; the
  values of $2\pi/(2k_F)$ are 24.1ML, 35.7ML and 52.6ML for $E_0=$ 20,
  10 and 5mRy, respectively. The peaks are much more violent for the
  minority-spin case (note the logarithmic scale there) because of the
  greater confinement due to stronger interface reflection.  }
\label{figGammaZnSe}
\end{figure}
\begin{figure}
\begin{center}
\resizebox{0.5\textwidth}{!}{
\includegraphics{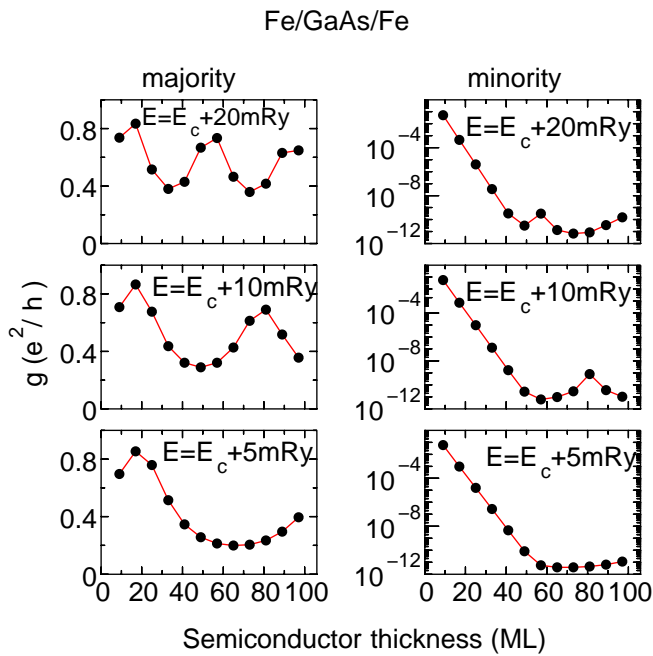}}
\end{center}
\caption{Same as in Fig.~\protect{\ref{figGammaZnSe}}, but for GaAs
  spacers. Here, the values of $2\pi/(2k_F)$ are 40ML, 64.5ML and
  95.2ML for $E_0=$ 20, 10 and 5mRy, respectively.  }
\label{figGammaGaAs}
\end{figure}

Another aspect of the matter is this: at the resonance values of
$k_zD$ we have also a formation of quantum well-like states in the
spacer. They are not bound, since the reflection is not total; the
``interactive'' change in the integrated density of states for each
$\mathbf{k}_{\parallel}$ because of them, compared to the bulk Fe,
is\cite{Bruno95}
\begin{equation}
\Delta N(E)=-\frac{1}{\pi}\mathrm{Im}\ln(1-|r_1| |r_2| e^{i(2k_z D+\phi_1+\phi_2)})
\label{eq3}
\end{equation}
per spin direction. Whenever such a quantum well state is met, a
resonance in the transmission probability is expected; the larger
$|r_1||r_2|$ is, the more peaked and localised in energy is the change
of the DOS and the transmission resonance.  \footnote{Actually, the
  maxima in the DOS do not always have to coincide with those in the
  transmission; a relative phase shift can occur, as pointed out in
  Ref.~\onlinecite{Taniguchi01}. However, such a shift is not expected
  in the present case if we follow the analysis of
  Ref.~\onlinecite{Taniguchi01}, since there are no zeros of the
  transmission.}

Dual to the oscillations of $g$ in $k$-space are oscillations in real
space, when the spacer thickness $D$ is varied while
$\mathbf{k}_{\parallel}$ is kept constant. As can be read out from
eq.~(\ref{eqres}), one expects a thickness period of $2\pi/(2k_z)$,
which for $\mathbf{k}_{\parallel}=0$ becomes $2\pi/(2k_F)$. Indeed, in
Figs.~\ref{figGammaZnSe} and \ref{figGammaGaAs} we can see this
oscillatory effect on the majority-spin conductance (left panels) for
both ZnSe and GaAs spacers, with exactly the predicted period. The
period gets longer for lower energy shifts, since they correspond to
lower $k_F$. On the other hand, larger $k_{\parallel}$ will result in
larger periods, until the limit value of $k_{\parallel}=k_F$; after
that $k_z$ becomes imaginary, and one has attenuation rather than
propagation of the wave, described by the complex band structure, as
in a tunnel junction.

Similarly, the minority-spin conductance oscillates with the same
period as seen in Figs.~\ref{figGammaZnSe} and \ref{figGammaGaAs}
(right pannels), but for the reasons mentioned before the peaks are
much more pronounced; note that in this case a logarithmic scale was
used for the intensities. It should be noted that there is, in
particular for GaAs, an initial exponential decrease in the
conductance, before the oscillations start, as can be seen from the
characteristic linear behaviour in the logarithmic scale. This
originates from decaying states with complex Bloch vectors, which
contribute to the conductance by tunneling. Indeed, minority-spin
states incident from Fe at $E_F$ having the $\Delta_5(C_{4v})$ and
$\Delta_2(C_{4v})$ symmetry (see Fig.~\ref{figFeGaAsZnSebands}) cannot
couple to the SC $\Delta_1(C_{2v})$ conduction band, but they can
couple to decaying SC states that have the correct symmetry. In this
way, if the thickness of the spacer is moderate, they can have an
important contribution to the current through
tunneling.\cite{Mavropoulos00} For larger thicknesses they become
unimportant, and the asymptotic oscillatory behaviour appears.  This
situation of co-existence of tunneling current with ``normal'' current
is much stronger in GaAs, because it has a smaller band gap than ZnSe,
and thus the decay length of such evanescent states is much longer.

\begin{figure}
\begin{center}
\resizebox{0.45\textwidth}{!}{
\includegraphics{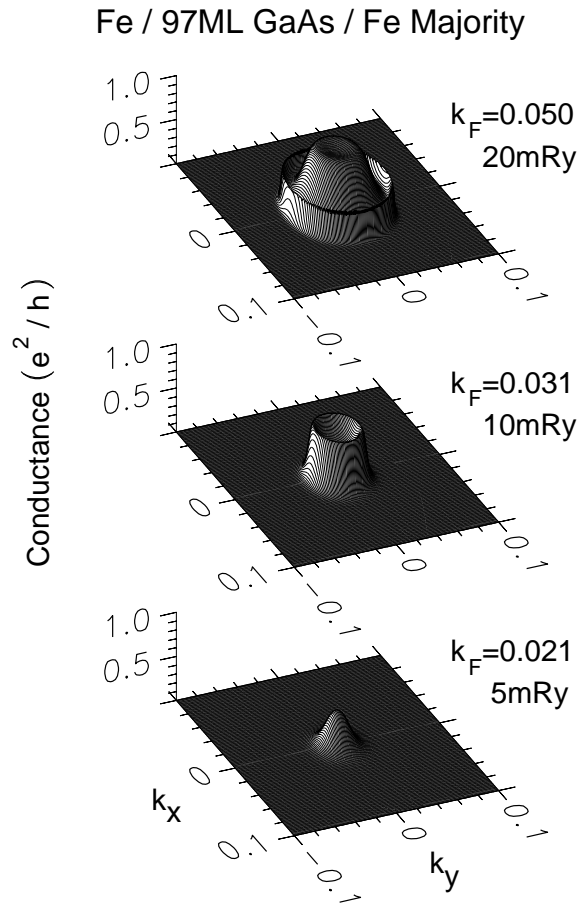}}
\end{center}
\caption{Conductance ($\mathbf{k}_{\parallel}$-resolved) of
  majority-spin electrons, in the case of a Fe/ 97ML GaAs /Fe
  junction. Top: $E=E_c+20\mbox{mRy}$, $k_F=0.050$; Middle:
  $E=E_c+10\mbox{mRy}$, $k_F=0.031$; Bottom: $E=E_c+5\mbox{mRy}$,
  $k_F=0.021$.  The $\mathbf{k}_{\parallel}$ axes are along the
  $\bar{\Gamma}-\bar{\mathrm{M}}$ directions.}
\label{figGaAs2dMaj}
\end{figure}
\begin{figure}
\begin{center}
\resizebox{0.45\textwidth}{!}{
\includegraphics{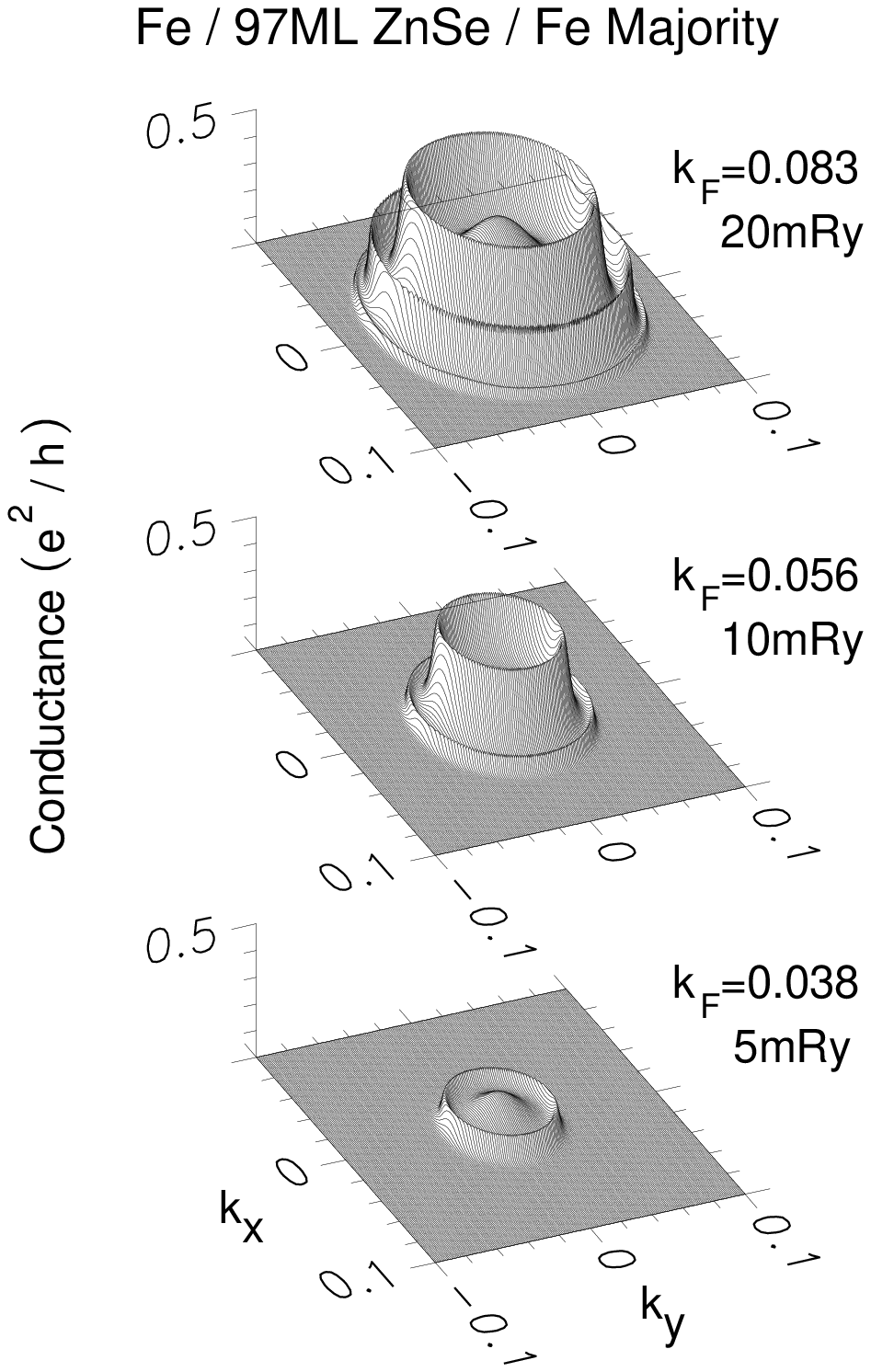}}
\end{center}
\caption{Conductance ($\mathbf{k}_{\parallel}$-resolved) of
  majority-spin electrons, in the case of a Fe/ 97ML ZnSe /Fe
  junction. Top: $E=E_c+20\mbox{mRy}$,
  $k_F=0.083$; Middle: $E=E_c+10\mbox{mRy}$,
  $k_F=0.056$; Bottom: $E=E_c+5\mbox{mRy}$,
  $k_F=0.038$.  The $\mathbf{k}_{\parallel}$
  axes are along the $\bar{\Gamma}-\bar{\mathrm{M}}$ directions.}
\label{figZnSe2dMaj}
\end{figure}
\begin{figure}
\begin{center}
\resizebox{0.45\textwidth}{!}{
\includegraphics{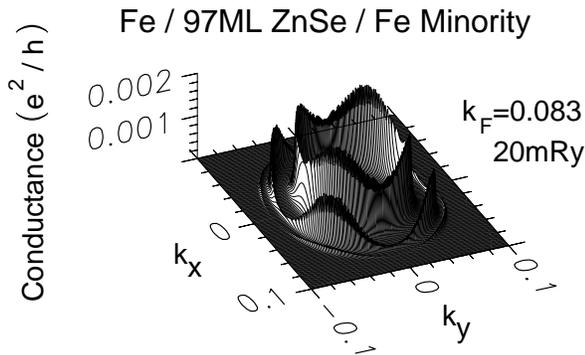}}
\end{center}
\caption{Conductance ($\mathbf{k}_{\parallel}$-resolved) of
  minority-spin electrons, in the case of a Fe/ 97ML ZnSe /Fe junction
  for $E=E_c+20\mbox{mRy}$, $k_F=0.083$; an octuple symmetry is
  evident. The $\mathbf{k}_{\parallel}$ axes are along the
  $\bar{\Gamma}-\bar{\mathrm{M}}$ directions.}
\label{fig2dZnSeMin}
\end{figure}
In Figs.~\ref{figGaAs2dMaj} and \ref{figZnSe2dMaj}, the majority-spin
$g(\mathbf{k}_{\parallel})$ is demonstrated for 97ML-thick spacers of
GaAs and ZnSe, for gate voltage shifts of $E_0=5$mRy, 10mRy and 20mRy.
The conductance resonances form rings around
$\mathbf{k}_{\parallel}=0$, up to $k_F$; they are what one expects by
rotating the graphs of Fig.~\ref{figZnSeGaAscond} around the origin.
It is remarkable that the majority-spin conductance is quite isotropic
in all cases. In contrast, we find the minority-spin conductance rings
to reflect more the quadruplicate structure of the surface Brillouin
zone, but seem actually (by inspection) to obey one extra symmetry
operation and to be octuple, as if the group were $C_{4v}$. This is
observed in all cases, and is mostly evident in the case of ZnSe with
$E_0=20$mRy, where $k_F$ is largest; this is shown in
Fig.~\ref{fig2dZnSeMin}.  We shall give the explanation of these
observations together with the analysis of similar data for the
antiparallel magnetic configuration of the leads, at the end of the
next section. Evidently, the majority-spin conductance retains its
dominance over its minority counterpart; the
$\mathbf{k}_{\parallel}$-integrated conductance is presented in
Table~\ref{table1}.

\section{Antiparallel moment in the leads - Magnetoresistance \label{sec7}}

From the analysis presented in Sections \ref{sec4} and \ref{sec5} one
should expect a strong reduction of the conductance if the magnetic
moments of the leads have an antiparallel orientation. If the moment
of, say, the second lead is reversed, then the majority and minority
bands will be interchanged there. So, the incoming minority-spin
electrons will be nearly blocked at the first interface, while the
incoming majority-spin electrons will propagate up to the second
interface but suffer almost total reflection there, since they will
encounter the states of the $\Delta_{2'}(C_{4v})$-type to which they
do not couple well. Again the situation is analogous to the one
encountered in the case of tunneling barriers.\cite{MacLaren99}

\begin{figure}
\begin{center}
\resizebox{0.5\textwidth}{!}{
\includegraphics{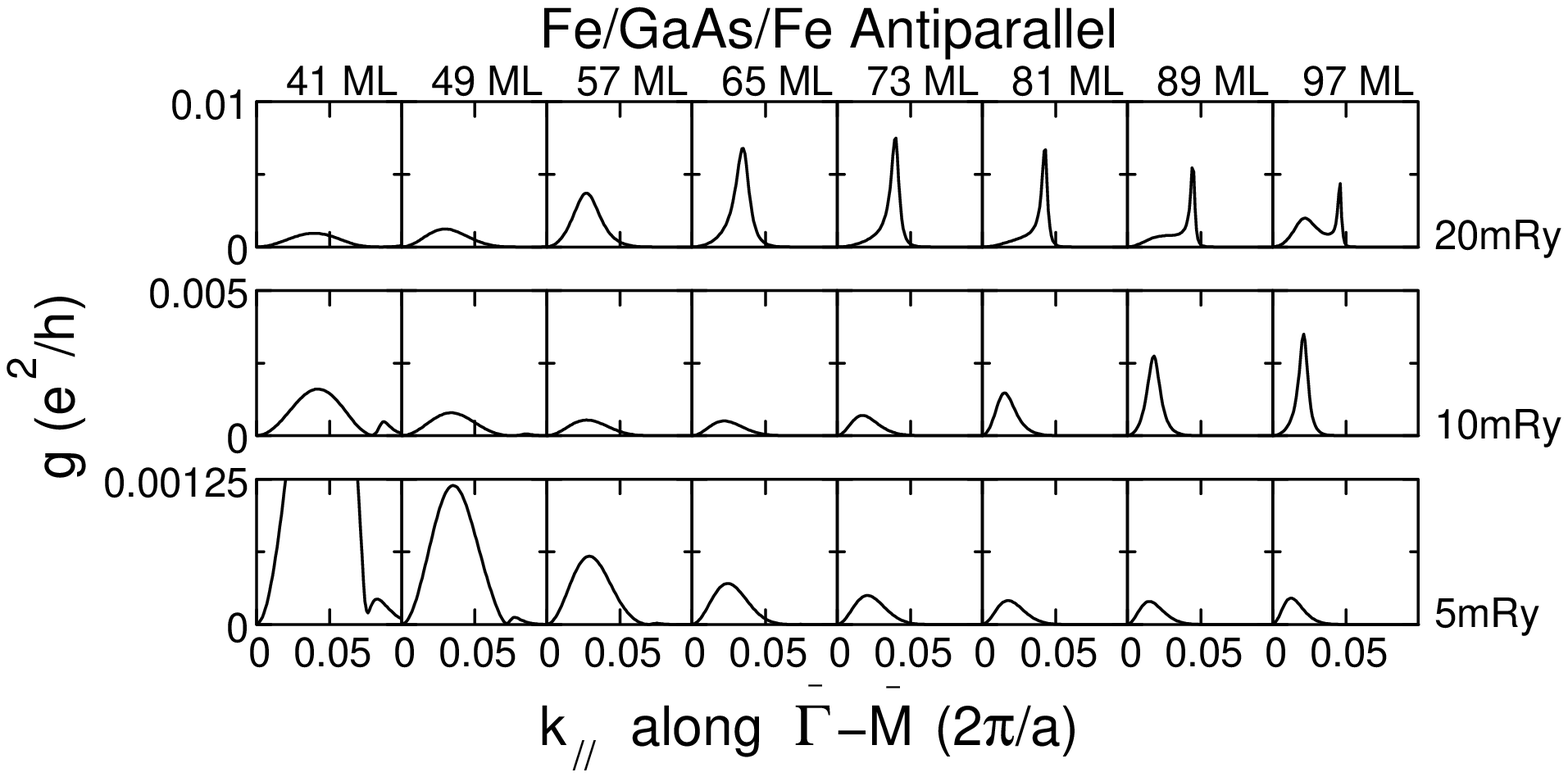}}
\resizebox{0.5\textwidth}{!}{
\includegraphics{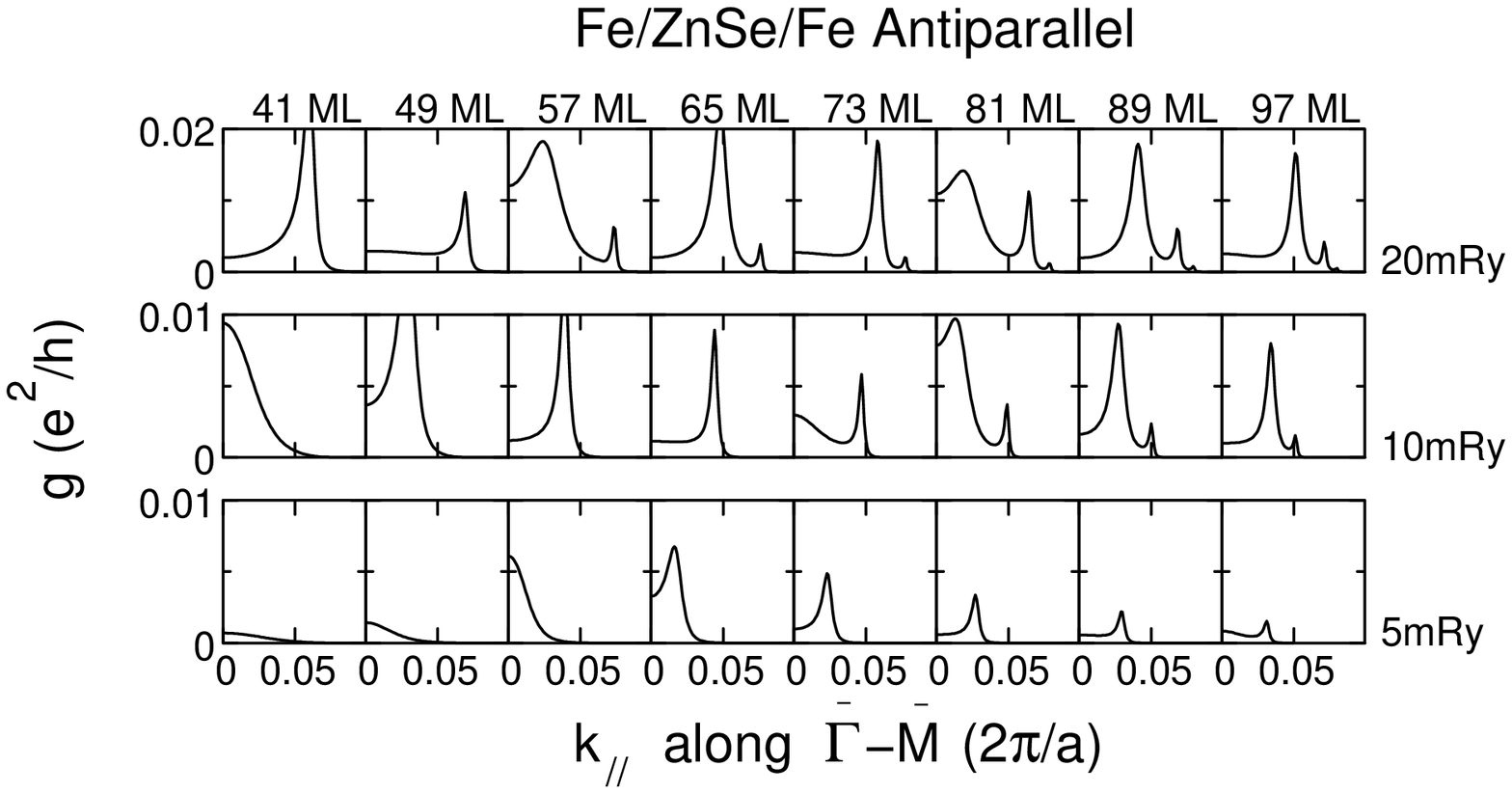}}
\end{center}
\caption{Fe/GaAs/Fe (top) and Fe/ZnSe/Fe (bottom) conductance per spin
  channel along $\bar{\Gamma}-\bar{\mathrm{M}}$ ($k_x$), for the
  antiparallel magnetic configuration of the leads, and for several
  spacer thicknesses. The values are much lower than the majority-spin
  conductance and much higher than the minority-spin conductance in
  the parallel case. The $k_F$-values are: 0.021, 0.031, and 0.050 for
  $E_0=5$, 10, and 20mRy, respectively.}
\label{figAllAF}
\end{figure}

Indeed, in Fig.~\ref{figAllAF} we see the that the conductance in the
antiparallel configuration is calculated to be orders of magnitude
lower than the majority-spin conductance (and the total one) of the
parallel configuration, but still orders of magnitude higher than the
minority-spin conductance of the parallel configuration. The effect
can be understood in terms of the reflectance and transmittance at the
interfaces. If $T_{\mathrm{si}}^{\uparrow}$ is the (high)
single-interface transmision probability involving majority Fe states
and $T_{\mathrm{si}}^{\downarrow}$ is the (low) one involving minority
Fe states, with $T_{\mathrm{si}}^{\uparrow}\gg
T_{\mathrm{si}}^{\downarrow}$, then in the case of parallel alignment
the majority electrons will have a total transmission probability from
both interfaces of the order of $T_{\mathrm{tot}}\sim
(T_{\mathrm{si}}^{\uparrow})^2$ (neglecting resonance effects), the
minority ones $(T_{\mathrm{si}}^{\downarrow})^2$, while the
antiparallel-configuration electrons will have
$T_{\mathrm{si}}^{\uparrow}T_{\mathrm{si}}^{\downarrow}$ for each spin
channel.  Evidently, $(T_{\mathrm{si}}^{\uparrow})^2\gg
T_{\mathrm{si}}^{\uparrow}T_{\mathrm{si}}^{\downarrow}\gg
(T_{\mathrm{si}}^{\downarrow})^2$, {\it q.e.d.}. We observe, by the
way, that this line of thought suggests that in the antiparallel
configuration the conductance $g_{\uparrow\downarrow}$ (per spin
channel) is the geometrical average of the conductances of the two
spin channels in the parallel case: $g_{\uparrow\downarrow}=
\sqrt{g_{\uparrow\uparrow}g_{\downarrow\downarrow}}$ (to be valid but
for backscattering effects). This is true for $\mathbf{k}_{\parallel}$
in certain directions of the surface Brillouin zone, {\it i.e.}~along
$k_x$ and $k_y$ (the cubic axes), including of course
$\mathbf{k}_{\parallel}=0$. At such $k$-points, the transmission
through the first interface (Fe into SC) is the same as through the
second (SC into Fe); however, for other
$\mathbf{k}_{\parallel}$-points this is not true, so spin-up and
spin-down electrons have different $g(\mathbf{k}_{\parallel})$ and
only equal $\mathbf{k}_{\parallel}$-integrated $g$ as shall be
explained in the end of the section. Our numerical results verify
this. So, for an arbitary $\mathbf{k}_{\parallel}$-point, the
geometric average relation can hold at most for the order of
magnitude. We note in passing that, if we had a spacer material with
$C_{4v}$ interface symmetry, as {\it e.g.}~MgO, the geometric average
rule would not hold at all, because the minority Fe
$\Delta_{2'}(C_{4v})$-state would be orthogonal to the spacer
$\Delta_1(C_{4v})$ conduction band; then the minority electrons would
reach the second interface only through a complex band with
exponentially damped probability and the assumptions of the
two-reflectance-argument would not hold.

\begin{figure}
\begin{center}
\resizebox{0.45\textwidth}{!}{
\includegraphics{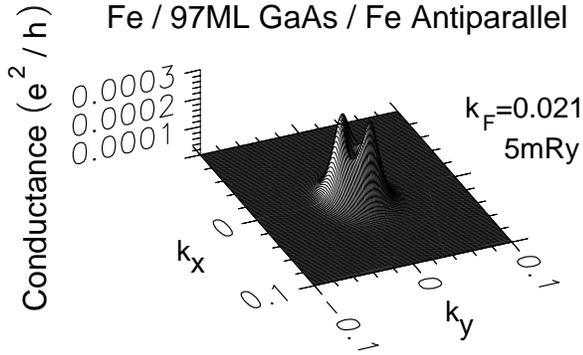}}
\end{center}
\caption{Conductance ($\mathbf{k}_{\parallel}$-resolved) of incoming
  majority electrons, in the case of a Fe/ 97ML GaAs /Fe junction with
  \emph{antiparallel} magnetic orientation of the two Fe leads, for
  $E=E_c+5\mbox{mRy}$, $k_F=0.021$; a quadruplicate symmetry is evident.
  The $\mathbf{k}_{\parallel}$ axes are along the
  $\bar{\Gamma}-\bar{\mathrm{M}}$ directions.}
\label{fig2dGaAsAF}
\end{figure}
For a large spacer thickness of 97MLs we see in Fig.~\ref{fig2dGaAsAF}
the $\mathbf{k}_{\parallel}$-resolved conductance for the transmission
from incoming majority-spin to outgoing minority-spin channels.  In
Table~\ref{table1}, we see the integrated (over the SBZ) conductance
for several gate voltage parameters $E_0$ in the case of 97ML-thick
spacers, together with the spin current polarisation
$P=(g_{\uparrow\uparrow}-g_{\downarrow\downarrow})/
(g_{\uparrow\uparrow}+g_{\downarrow\downarrow})$ and the
magnetoresistance (MR) ratio defined as
$(g_{\downarrow\uparrow}+g_{\uparrow\downarrow})/
(g_{\uparrow\uparrow}+g_{\downarrow\downarrow})$ (the so-called
``pessimistic definition'').
\begin{table*}
\caption{Calculated current polarisation and magnetoresistance
  (MR) ratio in the case of 97ML-thick spacers of ZnSe and GaAs for
  several gate voltage shifts $E_F-E_c$; both are close to the ideal
  100\%. The spin-dependent conductance $g$ integrated over the
  surface Brillouin zone is also shown for both cases of magnetic
  orientation of the leads: parallel (majority and minority) and
  antiparallel per spin (the same for the two spin channels).
  \label{table1}}
\begin{ruledtabular}
\begin{tabular}{cclcccc}
  Material&$E_F-E_c$ && {\small $g(e^2/h)$ (per unit-cell surface
    area)}& & Polarization & MR ratio\\ & & Majority & Minority &
  Antiparallel/spin & & \\ \hline \hline &5mRy (68meV) &$1.6
  \times10^{-7}$&$1.0 \times10^{-12}$&$2.6\times10^{-10}$
  &99.999\%&99.678\%\\ GaAs&10mRy (136meV)&$7.1 \times10^{-7}$&$2.1
  \times10^{-11}$&$2.7\times10^{-9}$ &99.994\%&99.229\%\\ &20mRy
  (272meV)&$1.9 \times10^{-6}$&$5.3 \times10^{-11}$&$7.8\times10^{-9}$
  &99.994\%&99.196\%\\ \hline &5mRy (68meV) &$1.7 \times10^{-7}$&$2.4
  \times10^{-12}$&$2.0\times10^{-9}$ &99.972\%&97.746\%\\ ZnSe&10mRy
  (136meV)&$8.2 \times10^{-7}$&$3.0 \times10^{-10}$&$1.4\times10^{-8}$
  &99.926\%&96.553\%\\ &20mRy (272meV)&$2.8 \times10^{-6}$&$2.4
  \times10^{-9}$ &$6.7\times10^{-8}$ &99.823\%&95.128\%\\ 
\end{tabular}
\end{ruledtabular}
\end{table*}
Evidently the calculated device acts as an almost ideal spin filter
and switch with extremely high MR ratio. For lower energy shifts the
spin filtering and MR ratio increase, because the allowed
$\mathbf{k}_{\parallel}$ close up to $\bar{\Gamma}$ and the states
have more and more $\Delta_1(C_{2v})$-character. Because of the
$\Delta_{2'}$ minority-spin state, however, the ideal 100\% cannot be
reached even in the limiting case; in contrast, it would be reached
{\it e.g.}~in the case of an MgO spacer because it exhibits
$C_{4v}$-symmetry.\footnote{The situation bears again an analogy with
  the magnetic tunnel junctions; in the limit of large spacer thickness, the
  current polarisation and MR ratio do not reach 100\% for
  ZnSe,\cite{MacLaren99} but they do reach it for
  MgO.\cite{Butler01,Mathon01}}

As promised at the end of the previous section, we now turn our
attention to the explanation of the circularly symmetric form of
$g(\mathbf{k}_{\parallel})$ for the majority electrons in the
parallel-alignment case, vs.~the octuple symmetry seen for the
minority electrons, and all this vs.~the quadruplicate symmetry in the
antiparallel-alignment case. As $\mathbf{k}_{\parallel}$ departs from
$\bar{\Gamma}$, the Fe and SC states do not belong exclusively to a
single representation any more, but are rather admixtures of the
various representations; but they still retain mostly the character
they had at $\bar{\Gamma}$. In the language of localised orbitals, the
majority-spin states are formed mostly by the circularly symmetric
$s+p_z+d_{z^2}$ orbitals (plus small admixtures away from
$\bar{\Gamma}$); the minority-spin states consist of $d_{xy}$ from the
$\Delta_{2'}(C_{4v})$-band, $p_x+p_y+d_{xz}+d_{yz}$ from the
$\Delta_5(C_{4v})$-band, and $d_{x^2-y^2}$ from the
$\Delta_2(C_{4v})$-band; finally the SC conduction band states consist
of $s+p_z+d_{xy}$ from the $\Delta_1(C_{2v})$-band.  Away from
$\bar{\Gamma}$, new orbitals start to contribute to each band, but in
amounts negligible for our discussion, since we remain close to
$\bar{\Gamma}$.

Firstly we concentrate on the coupling of the minority-spin states.
At exactly $\bar{\Gamma}$, the only combination that gives nonzero
inner product is $d_{xy}$ orbitals of Fe with $d_{xy}$-like states of
the SC; the rest of the combinations are inner products of symmetric
with antisymmetric wavefunctions, resulting to zero. As
$\mathbf{k}_{\parallel}$ departs from $\bar{\Gamma}$, the $p_x$,
$p_y$, $d_{xz}$, $d_{yz}$, and $d_{x^2-y^2}$ minority states of Fe
atoms neighbouring a particular SC atom at the interface obtain
slightly position-dependent phases as
$e^{i\mathbf{k}_{\parallel}\mathbf{r}}$; then the wavefunctions formed
by combining them obtain a small part symmetric around the SC atom,
and this gives nonzero inner product with the SC $s+p_z+d_{xy}$. This
overlap integral, in first approximation proportional to
$k_{\parallel}$, is different for the various directions of
$\mathbf{k}_{\parallel}$, following the pattern of $d_{xy}$. Clearly
then the bonding and the conductance must have a quadruplicate
symmetry in $\mathbf{k}_{\parallel}$-space, as does $d_{xy}$ in real
space.  By inspection of the Fig.~\ref{fig2dZnSeMin} we see an octuple
symmetry.  The explanation for the extra symmetry lies in the
zincblende geometry and the directionality of the bonding. Indeed, as
we enter the SC ({\it e.g.}~ZnSe) from the one lead, we encounter Zn
and then Se on the tetrahedral positions along the $(x,y)$ diagonal;
but as we leave it, we encounter Se and then Zn on the tetrahedral
positions along the $(x,-y)$-diagonal.  Thus, the directionality of
the SC $d_{xy}$-like states and consequently the bonding and
transmission properties of the two interfaces are equivalent but
rotated by $90^{\circ}$ to each other, so the combined transmission
obeys one extra symmetry operation and is octuple.

Secondly we focus on the coupling of the majority-spin states. There
the situation is simpler: Fe has only $s+p_z+d_{z^2}$ circularly
symmetric orbitals which can couple only to the SC $s+p_z$, but not to
$d_{xy}$. Thus no directionality is induced by the latter; even as we
depart from $\bar{\Gamma}$, the small difference in phase obtained by
neighbouring Fe sites gives only an antisymmetric part to the combined
wavefunction and this has still zero inner product with the SC
$s+p_z+d_{xy}$. The result is that the bonding and transmission
properties for majority are isotropic arround $\bar{\Gamma}$.

Finally we look at the antiparallel magnetic configuration of the
leads.  There, one either enters with circularly symmetric
transmission via majority and exits with the quadruplicate symmetry via
minority with a quadruplicate net result, as seen in
Fig.~\ref{fig2dGaAsAF}, or, for the opposite spin, enters with
quadruplicate symmetry via minority and exits with circularly symmetric
transmission via majority, again with a quadruplicate net result.  For the
two last cases, by the way, the $g(\mathbf{k}_{\parallel})$ are
rotated to each other by $90^{\circ}$, again due to the aforementioned
direction difference in the bonding; thus only along $k_x$ and $k_y$
is $g(\mathbf{k}_{\parallel})$ the same for the two spin directions in
the antiparallel case.

The same symmetry of $g(\mathbf{k}_{\parallel})$ as here is seen in
results for tunneling Fe/ZnSe/Fe junctions,\cite{MacLaren99} so once
more we see the formal connection between spin injection and
tunneling.


\section{Limitations and summary \label{sec8}}

Before summarising, we shall briefly discuss the limitations of our
approach and the relevance to realistic experimental situations.  Two
main points must be addressed here: (i) the influence of diffuse,
$\mathbf{k}_{\parallel}$-violating scattering and (ii) the possible
effect of a Schottky barrier.  As for point (i), it is true that the
formation of terraces or steps and interdiffusion lead to diffuse
scattering. To what extent this reduces the control over the
conductance must be examined seperately in each case and is a huge but
challenging task. In the case of the Fe/GaAs interface it is known
that in growing of Fe on GaAs the As atoms act like surfactants
forming always an As monolayer on Fe. This is of course an indication
that the interface structure is not perfect. But progress is being
done and one can reasonably hope that the quality of the interfaces
will increase a lot in the future.

About point (ii), Schottky barriers are known to extend over
mesoscopic lengths, especially when the doping is low. However,
techniques to use quantum well structures have resulted in lowering
the conducion band under the Fermi level without direct impurity
doping; such a situation would be modeled by our ``gate voltage''
parameter $E_0$ in addition to a real gate voltage. Then the Schottky
barrier would be much shorter, in fact being determined by the
Fermi level pinning due to the metal-induced gap states. On the other
hand, in the single-interface calculations for spin injection by
Wunnicke {\it et al.}\cite{Wunnicke02} the effect of a Schottky
barrier has been studied by emulating it with a long region near the
interface where the SC potentials were kept to their physical
unshifted positions, and the electrons had to really tunnel into the
conduction band. The result was quite encouraging, giving still an
extremely high current spin polarisation.

To summarise, we have performed {\it ab initio} calculations of the
spin-dependent transport through Fe/GaAs/Fe and Fe/ZnSe/Fe (001)
junctions, with a gate voltage parameter acting on the semiconductor
so that the Fermi level lies slightly in the conduction band. The
electron transport was supposed to be completely ballistic, assuming a
perfect interface structure and two-dimensional periodicity
perpendicular to the direction of growth. Under these assumptions we
have shown that such systems can exhibit an extremely high degree of
current spin polarisation and also a magnetoresistance ratio
approaching the ideal 100\%.  We have been able to trace down these
nice properties to the difference in the bulk band structure for the
two spin directions of Fe, and also to the difference in the bonding
of majority- and minority-spin states at the interface with the
semiconductor. In the same terms we have explained the high
magnetoresistance values. We have also examined interesting
interference effects that show up in such a junction due to the
presence of two, rather than one, interfaces, and discussed the
question whether these effects can invert the detected current
polarisation.

We have seen that the understanding of these systems stands in close
connection with the understanding of ballistic magnetic tunnel
junctions, if one formally replaces the band structure near the center
of the conduction band of the semiconductor with the complex band
structure in the gap region. In both cases, it is important that very
few states perform the conduction, namely the ones near the center of
the surface Brillouin zone; to know the properties of these states
means to have control over the conductance.

We have concluded that the control over the desired properties of such
systems is best when one deals with ballistic transport. Diffuse
scattering, particularly at the interface, would intermix the various
conducting channels and cause the injection efficiency and
magnetoresistance to drop; on the other hand, clean and abrupt
interfaces preserve $\mathbf{k}_{\parallel}$ and act as spin-selective
transmitters and detectors.

\begin{acknowledgments}
  The authors gratefully acknowledge support from the RT Network of
  {\em Computational Magnetoelectronics} (Contract No:
  RTN1-1999-00145) of the European Commission.
\end{acknowledgments}


\end{document}